\documentclass[preprint]{aastex63}

\usepackage{hyperref}

\received{April 28, 2020}
\revised{May 12, 2020}
\accepted{May 14, 2020}
\submitjournal{ApJ}

\shorttitle{Solar Plasma Heating by Flux Cancellation}
\shortauthors{Park et al.}

\graphicspath{{./}{figures/}}

\begin{document}

\title{An Observational Test of Solar Plasma Heating by Magnetic Flux Cancellation}

\correspondingauthor{Sung-Hong Park}
\email{shpark@isee.nagoya-u.ac.jp}

\author[0000-0001-9149-6547]{Sung-Hong Park}
\affiliation{Institute for Space-Earth Environmental Research, Nagoya University, Nagoya, Japan}

\begin{abstract}
Recent observations suggest that magnetic flux cancellation may play a crucial role in heating the Sun's upper atmosphere (chromosphere, transition region, corona). Here, we intended to validate an analytic model for magnetic reconnection and consequent coronal heating, driven by a pair of converging and cancelling magnetic flux sources of opposite polarities. For this test, we analyzed photospheric magnetic field and multi-wavelength UV/EUV observations of a small-scale flux cancellation event in a quiet-Sun internetwork region over a target interval of 5.2 hr. The observed cancellation event exhibits a converging motion of two opposite-polarity magnetic patches on the photosphere and red-shifted Doppler velocities (downflows) therein consistently over the target interval, with a decrease in magnetic flux of both polarities at a rate of 10$^{15}$\,Mx\,s$^{-1}$. Several impulsive EUV brightenings, with differential emission measure values peaked at 1.6\,--\,2.0\,MK, are also observed in the shape of arcades with their two footpoints anchored in the two patches. The rate of magnetic energy released as heat at the flux cancellation region is estimated to be in the range of (0.2\,--\,1)$\times$10$^{24}$\,erg\,s$^{-1}$ over the target interval, which can satisfy the requirement of previously reported heating rates for the quiet-Sun corona. Finally, both short-term (a few to several tens of minutes) variations and long-term (a few hours) trends in the magnetic energy release rate are clearly shown in the estimated rate of radiative energy loss of electrons at temperatures above 2.0\,MK. All these observational findings support the validity of the investigated reconnection model for plasma heating in the upper solar atmosphere by flux cancellation.
\end{abstract}


\section{Introduction} \label{sec:intro}
Reconnection is a physical process in a magnetized plasma, in which magnetic field lines with antiparallel components are brought together in a current sheet or at a magnetic null point, break up, and then reconnect to establish a new, lower-energy configuration \citep{2000mare.book.....P}. Through this reconnection process, free energy stored in stressed magnetic fields is liberated in the forms of plasma thermal and kinetic energy, and hence radiation, all of which may be relevant for solar activity including explosive, transient and/or eruptive events. In order to understand how the reconnection process facilitates these solar eruptive events, there have been many recent developments in modelling reconnection in kinetic, hybrid and magnetohydrodynamic (MHD) frameworks with a variety of different current sheet structures in two or three dimensions \citep[for reviews, see][and references therein]{1995GMS....90..139L,1999SoPh..190....1P,2010RvMP...82..603Y,2011AdSpR..47.1508P,2012RSPTA.370.3169P,2016RSPSA.47260479Z}. In addition, many observed signatures have been reported, mainly interpreting large-scale morphological structures, plasma motions, heating or acceleration as signatures suggested in reconnection models.  Examples include hot, cusp-shaped flare loops in soft X-ray \citep{1992PASJ...44L..63T}, loop-top hard X-ray sources \citep{1994Natur.371..495M,2003ApJ...596L.251S}, current sheet-like structures \citep{2003JGRA..108.1440W,2010ApJ...723L..28L,2016NatCo...711837X,2018ApJ...853L..18Y}, plasma inflows \citep{2001ApJ...546L..69Y,2009ApJ...703..877L,2013NatPh...9..489S,2015NatCo...6.7598S,2015ApJ...798L..11Y} and outflows in the vicinity of current sheet-like structures or coronal magnetic null points \citep{2004ASPC..325..361A,2010ApJ...722..329S,2013ApJ...767..168L,2016ApJ...818L..27C}, hot plasmoids ejected from flare sites \citep{1995ApJ...451L..83S,2010ApJ...711.1062N,2012ApJ...745L...6T}, inverse Y-shaped jet-like features \citep{1999ApJ...513L..75C,2007Sci...318.1591S,2014Sci...346A.315T}, interaction between coronal loops \citep{1996SSRv...77....1S,2006ApJ...646..605S,2018ApJ...854..178N}, and drifting pulsating structures observed in radio dynamic spectra \citep{2000A&A...360..715K,2007SoPh..241...77N}. However, more direct observational evidence that quantitatively characterizes magnetic reconnection in the solar atmosphere, particularly in the context of energy release, has been rarely reported.

Photospheric magnetic flux cancellation has been thought to appear in consequence of (or as a cause of) magnetic reconnection; it is observed as the mutual disappearance of converging photospheric magnetic patches of opposite polarities \citep{1985AuJPh..38..855L,1985AuJPh..38..929M}. Two scenarios are often employed, the emergence of U-shaped loops and the submergence of $\Omega$-shaped loops \citep{1987ARA&A..25...83Z,2007ApJ...671..990K}. In these scenarios, magnetic patches of opposite polarities are brought closer to each other, make a field-line connection via the process of magnetic reconnection \citep{1993SoPh..143..119W}, and then the newly formed U-shaped/$\Omega$-shaped loops continue to emerge/submerge respectively. Such convergence and gradual disappearance of the opposite-polarity magnetic patches are found in photospheric magnetogram time sequences, sometimes accompanied by consistently blue-shifted or red-shifted Doppler velocities, on average, within the patches in the course of cancellation when observed near the solar disk center \citep{2009RAA.....9..921Z,2010ApJ...713..325I,2019A&A...622A.200K}. Flux cancellation events are also frequently observed with transient brightenings in many spectral lines in the vicinity of cancellation regions \citep{2003PASJ...55..313Y,2009ApJ...700L.145V,2013ApJ...763...97W,2016MNRAS.463.2190N,2016ApJ...823..110R,2016A&A...596A..15A}.

Recent observations of solar magnetic fields with unprecedented spatial and temporal resolutions have revealed the presence of a complex, mixed polarity magnetic field distribution, as well as a significant, rapid decrease of magnetic flux, at the photospheric base of active region coronal loops as well as in both quiet-Sun internetwork and network regions \citep{2009ApJ...704L..71P,2014ApJ...795L..24T,2017ApJS..229...17S,2018ApJ...853L..26H,2018ApJ...861..135Y,2018A&A...615L...9C,2019ApJ...873...75S,2019ApJ...887...56T}. For example, \citet{2017ApJS..229....4C} detected the cancellation of photospheric magnetic flux at a rate of 10$^{15}$\,Mx\,s$^{-1}$ at the footpoints of bright coronal loops in extreme ultraviolet (EUV), using line-of-sight (LOS) magnetic field data from the Imaging Magnetograph eXperiment \citep[IMaX;][]{2011SoPh..268...57M} on the SUNRISE balloon-borne observatory \citep{2010ApJ...723L.127S,2017ApJS..229....2S}. At the loop footpoints, they also found small-scale chromospheric jets during the flux cancellation which may supply mass and energy to overlying coronal loops.

Motivated by such high-resolution observations of magnetic flux cancellation, \citet{2018ApJ...862L..24P} proposed an analytic model of magnetic reconnection for solar chromospheric and coronal heating. The model allows for calculating the rate of magnetic energy converted to heat at a reconnecting current sheet, based on a few key parameters which may be derived from photospheric magnetic field observations. In order to examine the model validity, two-dimensional (2D) and three-dimensional (3D) resistive MHD simulations of two cancelling magnetic sources of opposite polarities in the presence of an overlying, background horizontal magnetic field and a stratified atmosphere have been constructed \citep{2019ApJ...872...32S,2020ApJ...891...52S}. It has been found from these simulations that the analytic model's estimate for the magnetic energy release rate is in good agreement with that explicitly calculated from the values assigned to the associated simulation parameters. 

In this study we analyze photospheric magnetic field and multi-wavelength UV/EUV observations of a small-scale flux cancellation event in a quiet-Sun internetwork region. We intend to (1) estimate the rate of magnetic energy released as heat as proposed in the \citet{2018ApJ...862L..24P} reconnection model and (2) quantitatively validate how the model performs comparing the estimated magnetic energy release rate with the rate of energy loss by radiation. The model is described in \S\,\ref{sec:model}, the data analysis techniques are presented in \S\,\ref{sec:data_analysis}, and analysis results in \S\,\ref{sec:results}. Finally, in \S\,\ref{sec:discussion}, we summarize our main findings.

\begin{figure}[h!]
\centering
\includegraphics[width=0.75\textwidth]{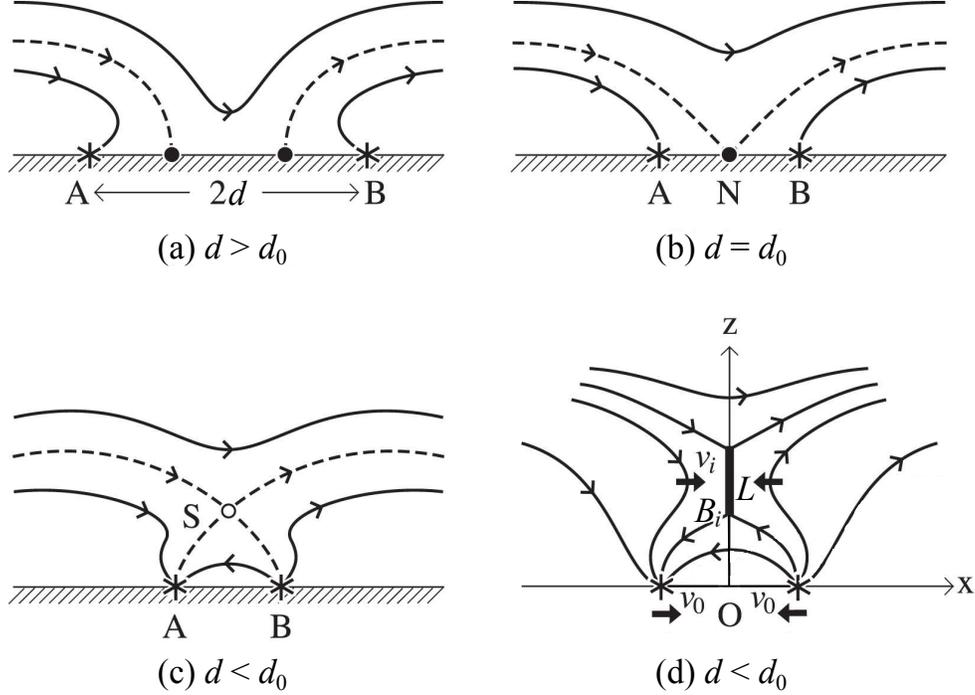}
\caption{Reproduction of Figure 3 of \citet{2018ApJ...862L..24P}, showing a cartoon of magnetic field structures for two opposite-polarity sources (A and B) of magnetic flux $\pm F$ situated in an overlying uniform horizontal magnetic field $B_{0}$ when (a) $d$\,$>$\,$d_{0}$, (b) $d$\,$=$\,$d_{0}$, and (c) $d$\,$<$\,$d_{0}$, where $d$ is the half-separation between the two flux sources and $d_{0}$ is the flux interaction distance. Separatrix magnetic field lines are marked by dashed lines, other magnetic field lines by solid lines, a null point (N) by a filled circle, and a separator (S) by an open circle.}
\label{fig:priest_model}
\end{figure}

\section{Reconnection Model for Heating} \label{sec:model}
An analytic model for magnetic reconnection driven by converging photospheric magnetic patches of opposite polarities was proposed by \citet{2018ApJ...862L..24P}, in which two opposite-polarity photospheric sources (A and B as marked in Figure~\ref{fig:priest_model}) with magnetic flux $\pm F$ situated in an overlying uniform horizontal magnetic field $B_{0}$. The two sources are initially separated from each other by a distance $d$, and start to reconnect as they approach each other. Figure~\ref{fig:priest_model} schematically describes (a) the initial configuration of the model magnetic field lines and separatrix magnetic field lines, and (b--c) their reconfiguration over the course of reconnection. The configuration of the model magnetic field can be described by the value of a key parameter, called the flux interaction distance $d_{0}$ \citep{1998ApJ...507..433L}, which for 3D sources, is written as 
\begin{equation}
d_{0} = \sqrt{\frac{F}{\pi B_{0}}}.
\label{eq:d0}
\end{equation}
When $d$\,$>$\,$d_{0}$ as shown in Figure~\ref{fig:priest_model}{a}, there is no magnetic field connecting the two sources and two first-order null points lie on the photosphere between the sources. In the case of $d$\,$=$\,$d_{0}$, there is a local bifurcation in which the nulls combine to form a high-order null at the origin (see Figure~\ref{fig:priest_model}{b}). Following \citet{2018ApJ...862L..24P}, as the sources approach closer to each other, such that $d$\,$<$\,$d_{0}$ (Figure~\ref{fig:priest_model}{c}), reconnection is driven and the location of a semicircular separator rises in the upper atmosphere to the height $z_{s}$ given by
\begin{equation}
z_{s} = \sqrt{{d_{}}^{2/3}\,{d_{0}}^{4/3} - {d_{}}^{2}}.
\label{eq:zs}
\end{equation}
In the case of fast reconnection, the total rate of magnetic energy released as heat is 
\begin{equation}
\frac{\mathrm{d}W}{\mathrm{d}t} = 0.4\,S_{i} = 0.8\,\frac{v_{i}{B_{i}}^{2}}{\mu} L L_{s} = \frac{1.6\pi}{3} \frac{v_{0}{B_{0}}^{2}}{\mu} {d_{0}}^{2} \frac{M_{A0}}{\alpha} \frac{1-(d/d_{0})^{4/3}}{(d/d_{0})^{2/3}},
\label{eq:dwdt}
\end{equation}
where $S_{i}$ is the Poynting flux flowing into both sides of the current sheet region of given length $L$ and depth $L_{s}$\,$=$\,$\pi z_{S}$ at the separator. $v_{i}$ and $B_{i}$ are the reconnection inflow speed and the strength of the magnetic field drawn into the current sheet, respectively. $\mathrm{d}W/\mathrm{d}t$ is derived based on the assumption that 40\% of the Poynting influx is converted to heat during fast reconnection \citep{2014masu.book.....P}. The physical quantities of $v_{i}$, $B_{i}$, $L$ and $L_{s}$, which are associated with the reconnection current sheet and magnetized plasma therein, can be basically derived as functions of $d$, $d_{0}$, $B_{0}$ and the flux source speed on the photosphere $v_{0}$\,$=$\,$\dot{d}$. An Alfv\'en Mach number $\alpha$ is defined as $v_{i}/v_{Ai}$ (where $v_{Ai}$\,$=$\,$B_{i}/\sqrt{\mu \rho_{i}}$ and $\rho_{i}$ is the density of the inflowing plasma), and a hybrid Alfv\'en Mach number $M_{A0}$ as $v_{0}/v_{A0}$ (where $v_{A0}$\,$=$\,$B_{0}/\sqrt{\mu \rho_{i}}$). To estimate $\mathrm{d}W/\mathrm{d}t$ for a flux cancellation event in a quiet-Sun region, we adopt here values of $\alpha$\,$=$\,0.1 and $M_{A0}$\,$=$\,0.1 \citep{2014masu.book.....P,2018ApJ...862L..24P}.

\section{Observations and Data Analysis} \label{sec:data_analysis}
An explosive, small-scale flux cancellation event was observed on 2017 November 7 in a quiet-Sun internetwork region (located at heliographic coordinates N05{\degr}E30{\degr}) by the \textit{Solar Dynamics Observatory} \citep[\textit{SDO},][]{2012SoPh..275....3P} and the \textit{Interface Region Imaging Spectrograph} \citep[\textit{IRIS},][]{2014SoPh..289.2733D}. In this study, we analyze the following dataset over an interval of 5.2 hr from 2017 November 07 00:00 TAI (hereafter referred to as the ``target interval''), comprising: (1) photospheric line-of-sight magnetograms with a spatial resolution of 1{\arcsec} (pixel size of 0.5{\arcsec}) obtained with a 45 s cadence by the \textit{SDO}/Helioseismic and Magnetic Imager \citep[HMI,][]{2012SoPh..275..207S}, (2) multi-wavelength solar images (with 0.6{\arcsec}/pixel) at 12 s cadence in seven EUV channels and at 24 s cadence in one UV channel by the \textit{SDO}/Atmospheric Imaging Assembly \citep[AIA,][]{2012SoPh..275...17L}, and (3) \textit{IRIS} slit-jaw images (SJIs) at the Si IV 1400\,{\AA} with a 40\,{\AA} bandpass at 36 s cadence. Note that \textit{IRIS} SJIs at 1400\,{\AA} are only available from 04:50 TAI on 2017 November 07. The observation dataset is summarized in Table~\ref{tbl:obs_data}.

\begin{deluxetable}{l|l|l}[b!]
\tablecaption{Data and Diagnostics}
\label{tbl:obs_data}
\tablehead{
\colhead{Source} & \colhead{Observable} & \colhead{Diagnostics}} 
\startdata
\textit{SDO}/HMI & Line-of-sight magnetograms & Photospheric magnetic flux \\
\textit{SDO}/AIA & EUV images at 94, 131, 171, 193, 211, 304 and 335\,{\AA} & $\log_{10}$\,T\,$=$\,$4.9 \-- 6.85$\,[K]\\
\textit{SDO}/AIA & UV images at 1600\,{\AA} & $\log_{10}$\,T\,$=$\,4\,[K]\\
\textit{IRIS} & Slit-jaw images at 1400\,{\AA} & $\log_{10}$\,T\,$=$\,$3.7 \-- 5.2 $\,[K]\\
\enddata
\end{deluxetable}

The radial component $B_\mathrm{r}$ of the magnetic field to the solar surface is derived from the LOS component $B_\mathrm{LOS}$, assuming the magnetic field is radial at the photosphere: i.e., $B_\mathrm{r}$\,=\,$B_\mathrm{LOS}$/cos($\theta$), where $\theta$ is the heliocentric angle of the given observed location (i.e., the angle between the LOS and the local normal to the surface) \--\ here $\cos(\theta)$\,$=$\,0.85\,--\,0.89. This assumption is consistent with plage field structure as observed by HMI, and more appropriate than other correction options given the nature of the field and the location \citep[refer to][]{2017SoPh..292...36L}.

\begin{figure}[t!]
\centering
\includegraphics[width=0.98\textwidth]{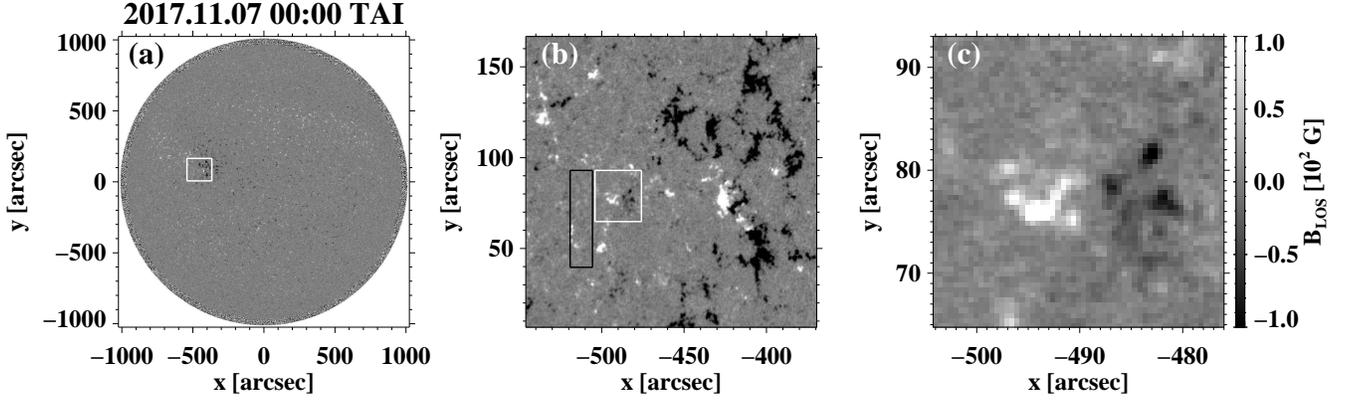}
\caption{A full-disk photospheric LOS magnetogram observed by \textit{SDO}/HMI on 2017 November 7 00:00 TAI is shown in panel (a). The white box in panel (a) indicates a quiet-Sun internetwork region shown in panel (b). Our region of interest (ROI, panel (c)) is marked with the white box in panel (b), in which a pair of opposite-polarity magnetic flux patches show a converging motion as well as a decrease in magnetic flux of both polarities (i.e., a so-called ``photospheric magnetic flux cancellation event''). The black box in panel (b) is a weaker-field region compared to the nearby ROI, which is chosen to calculate the average strength of the overlying horizontal magnetic field.}
\label{fig:target_region}
\end{figure}

As shown in Figure~\ref{fig:target_region}, the flux cancellation event is located in a quiet-Sun internetwork region denoted by the white box in the full-disk LOS magnetogram in panel (a). The white box in panel (b) outlines the region of interest (ROI, panel (c)) in which a pair of opposite-polarity and similar-sized magnetic flux patches in close proximity converge steadily over the target interval. The patches eventually show a concurrant decrease in magnetic flux in both the positive and negative polarities for the last 1.5 hr of the target interval. During the cancellation event, we find transient brightenings with lifetimes of a few to several tens of minutes from all UV and EUV wavelength channels on top of a long-term (a few hours) gradual increase and then decrease of the average UV/EUV intensities. A detailed description of the flux cancellation event is presented in Section\,\ref{sec:results}. 

In order to determine the parameters of $d$, $d_{0}$ and $v_{0}$ in Equation~(\ref{eq:dwdt}) for the pair of opposite-polarity magnetic patches in the ROI, using a sequence of co-aligned $B_\mathrm{r}$ images, we develop an algorithm for automatic identification and tracking of magnetic patches. First, patch identification is carried out for a given $B_\mathrm{r}$ image of the ROI at a single point in time, based on the following steps: (1) making a bitmap based on a given threshold value $B_\mathrm{th}$ of $|B_\mathrm{r}|$, containing one of three values, i.e., $+$1 for $B_\mathrm{r}$\,$\geq$\,$B_\mathrm{th}$, 0 for $-B_\mathrm{th}$\,$<$\,$B_\mathrm{r}$\,$<$\,$B_\mathrm{th}$, and $-$1 for $B_\mathrm{r}$\,$\leq$\,$-B_\mathrm{th}$, (2) finding and labeling positive polarity patches, each of which is defined as a group of pixels signed with $+$1 and also located next to each other; repeating the same for negative polarity patches (i.e., signed with $-$1), (3) grouping patches if the shortest distance between the patches is less than a grouping distance $d_\mathrm{gp}$, and (4) removing too small-sized patches (i.e., the total number of pixels belonging to a given patch\,$<$\,$N_\mathrm{min}$) and patch pixels located within a distance $d_\mathrm{sd}$ from all four sides of the image, if any. 
Next, all patches identified and labeled therein are tracked in time from two consecutive $B_\mathrm{r}$ images, considering four possible cases of moving, merging, splitting and newly emerging patches as follows. (1) Moving: if any single pixel of a patch $\mathcal{A}$ at $\mathrm{T_{0}}$ is matched with pixels of a patch $\mathcal{B}$ at $\mathrm{T_{1}}$\,=\,$\mathrm{T_{0}}$\,$+$\,45 s with the same polarity as $\mathcal{A}$, then $\mathcal{A}$ and $\mathcal{B}$ are considered as the same labeled patch. Note that when comparing the pixels between $\mathcal{A}$ and $\mathcal{B}$, a parameter $d_\mathrm{ds}$ for the maximum possible displacement in both $\pm$x and $\pm$y directions is used to take into account patch motions over 45 s. (2) Merging: if pixels of a patch $\mathcal{C}$ at $\mathrm{T_{1}}$ are matched with pixels of two or more patches at $\mathrm{T_{0}}$ even considering their motions by $\pm d_\mathrm{ds}$, then those patches at $\mathrm{T_{0}}$ are regarded as merging into one patch $\mathcal{C}$. (3) Splitting: if pixels of a patch $\mathcal{D}$ at $\mathrm{T_{0}}$ are matched with pixels of two or more patches (apart from each other with a separation distance greater than $d_\mathrm{sp}$), then we decide that $\mathcal{D}$ splits into two or more. The largest patch at $\mathrm{T_{1}}$ continues to be labeled $\mathcal{D}$ while the smaller patches are newly labeled at $\mathrm{T_{1}}$. (4) Newly emerging: if any single pixel of a patch $\mathcal{E}$ at $\mathrm{T_{1}}$ is not matched with pixels of all patches identified at $\mathrm{T_{0}}$ as well as their displacement by $\pm d_\mathrm{ds}$, then $\mathcal{E}$ is considered as a newly emerging patch. Here, values for all input parameters of the patch identification and tracking algorithm are set as follows: $B_\mathrm{th}$\,=\,20\,G, $d_\mathrm{gp}$\,=\,1.5{\arcsec}, $N_\mathrm{min}$\,=\,5 pixels, $d_\mathrm{ed}$\,=\,2.5{\arcsec}, $d_\mathrm{ds}$\,=\,0.5{\arcsec}, and $d_\mathrm{sp}$\,=\,4{\arcsec}. After an across-the-board survey with different sets of the input parameter values, the above values have been chosen in two respects: first, both the main positive and negative patches (as shown in Figure~\ref{fig:target_region}(a)) of the flux cancellation event are persistently tracked, and second, the center-of-mass (CoM) separation between the two patches (i.e., the distance between the flux-weighted centroid positions of the two patches) smoothly changes with a minimum of fluctuation over the target interval. 

\begin{figure}[t!]
\centering
\includegraphics[width=1\textwidth]{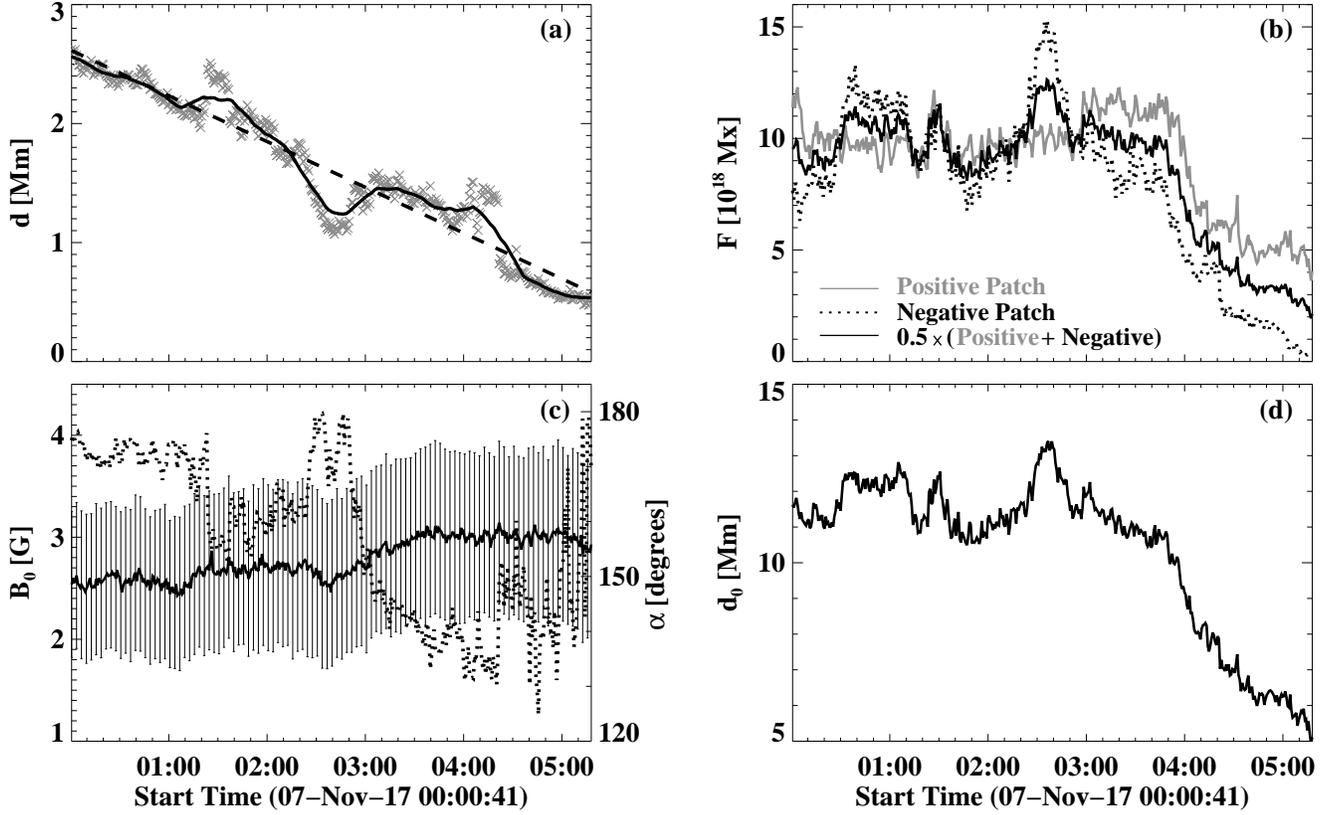}
\caption{Parameters of (a) CoM separation $d$ between the main positive and negative magnetic patches on the photosphere, (b) magnetic flux $F$, (c) average overlying horizontal magnetic field strength $B_{0}$ (solid line) with its standard deviation $\sigma(B_{0})$ (error bars) and (d) flux interaction distance $d_{0}$, derived from a sequence of co-aligned $B_\mathrm{r}$ images with the magnetic patch identification and tracking algorithm developed in this study. In panel (a), the solid and dashed lines represent, respectively, the 40-point (corresponding to 0.5 hr) running average of $d$ (raw data marked by cross symbols) and the least squares regression line of the running average. In panel (c), the dotted line indicates the angle $\alpha$ between the average overlying horizontal field vector and the line connecting the CoM of the positive patch to that of the negative patch.}
\label{fig:patch_outputs}
\end{figure}

With the outputs of the algorithm for identification and tracking of magnetic patches (i.e., pixel positions of tracked patches), we determine the CoM separation (Figure~\ref{fig:patch_outputs}{(a)}) between the two main patches of opposite polarities, as a function of time, which is assigned as $d$ in Equation~\ref{eq:dwdt}. The converging speed $v_{0}$ of $\sim$0.1 km/s is then derived from the slope of the least squares regression line of the 40-point (i.e., 0.5 hr) running average of $d$ over the target interval, in order to avoid some unreliable, sudden, large fluctuations mainly due to the fact that each of the main positive and negative patches occasionally but instantaneously merges with nearby patches and/or splits into smaller patches over the course of tracking. The adjusted R-squared value, widely used goodness-of-fit measure, is 0.95, indicating that the linear least squares regression method provides an adequate fit to the time series of $d$. The magnetic flux $F$ (which is the same between the positive and negative polarities in the proposed model by \citet{2018ApJ...862L..24P}) is defined here by the total unsigned flux of the two main positive and negative patches divided by a factor of 2 (see Figure~\ref{fig:patch_outputs}{(b)}). To estimate the overlying horizontal magnetic field strength ($B_{0}$, solid line in Figure~\ref{fig:patch_outputs}{(c)}), we first construct a potential magnetic field from a given photospheric $B_\mathrm{r}$ image of the full field of view in panel (b) of Figure~\ref{fig:target_region}. $B_{0}$ is then computed by averaging the horizontal component of the potential field, at heights ranging from 7 to 15\,Mm, over a weaker-field region (marked with the black box in Figure~\ref{fig:target_region}{(b)}) compared to the ROI. The weaker-field region considered here to calculate $B_{0}$ was selected to avoid the closed coronal magnetic field connecting the opposite-polarity patches, but to include a highly inclined field (on average, $>$\,75{\degr}) with respect to the radial direction. The specified height range of 7\,--\,15\,Mm was chosen, of which the lower limit is defined as the maximum height of $z_{s}$ over the target interval (see Figure~\ref{fig:time_profiles}{(b)}) and the upper limit as the height where the average strength of the horizontal potential field becomes smaller than that of the vertical field. The selected weaker-field region is found to have the following characteristics: (1) it is located close to the ROI; (2) the horizontal potential field in the selected region is, on average, stronger than the vertical field by a factor of $\sim$2 at heights of 7\,--\,15\,Mm; (3) the average angle of the overlying horizontal potential field vector relative to the line connecting the CoM of the positive patch to that of the negative patch ranges from 130 to 180{\degr} (refer to Figure~\ref{fig:patch_outputs}{(c)}). The flux interaction distance $d_{0}$ (plotted in Figure~\ref{fig:patch_outputs}{(d)}) is calculated from $F$ and $B_{0}$ as defined in Equation~\ref{eq:d0}.

The differential emission measure (DEM) is an estimate of the total number of electrons squared along the observed LOS (similar to a column mass) at a given temperature. It has been extensively used as a diagnostic tool to characterize temporal variations and spatial distributions of electron density and temperature in the solar atmosphere. The regularized inversion code developed by \citet{2012A&A...539A.146H} is used here to produce DEM maps at 48 s cadence from \textit{SDO}/AIA co-aligned images in six EUV channels centered at 94, 131, 171, 193, 211 and 335\,{\AA}, which have strong responses to logarithmic temperatures of $\log_{10}$\,T\,$=$\,6.85, 5.75, 5.95, 6.20, 6.25 and 5.35\,[K], respectively, from the latest Version 9 \citep{2019ApJS..241...22D} of CHIANTI. The temperature bins, as the inputs to the regularization algorithm, are chosen to be a total of 24 bins in the range of $\log_{10}$\,T\,$=$\,5.3\,--\,7.7\,[K], equally spaced on a logarithmic scale. In \citet{2012A&A...539A.146H}, it was found that this regularized inversion method is able to successfully recover the expected DEM from simulated data of a variety of model DEMs. Note that the inversion code provides uncertainties in both the DEM and temperatures, which allows us to estimate the accuracy of the regularized DEM solution.

\section{Results} \label{sec:results}
\begin{figure}[t!]
\centering
\includegraphics[width=1\textwidth]{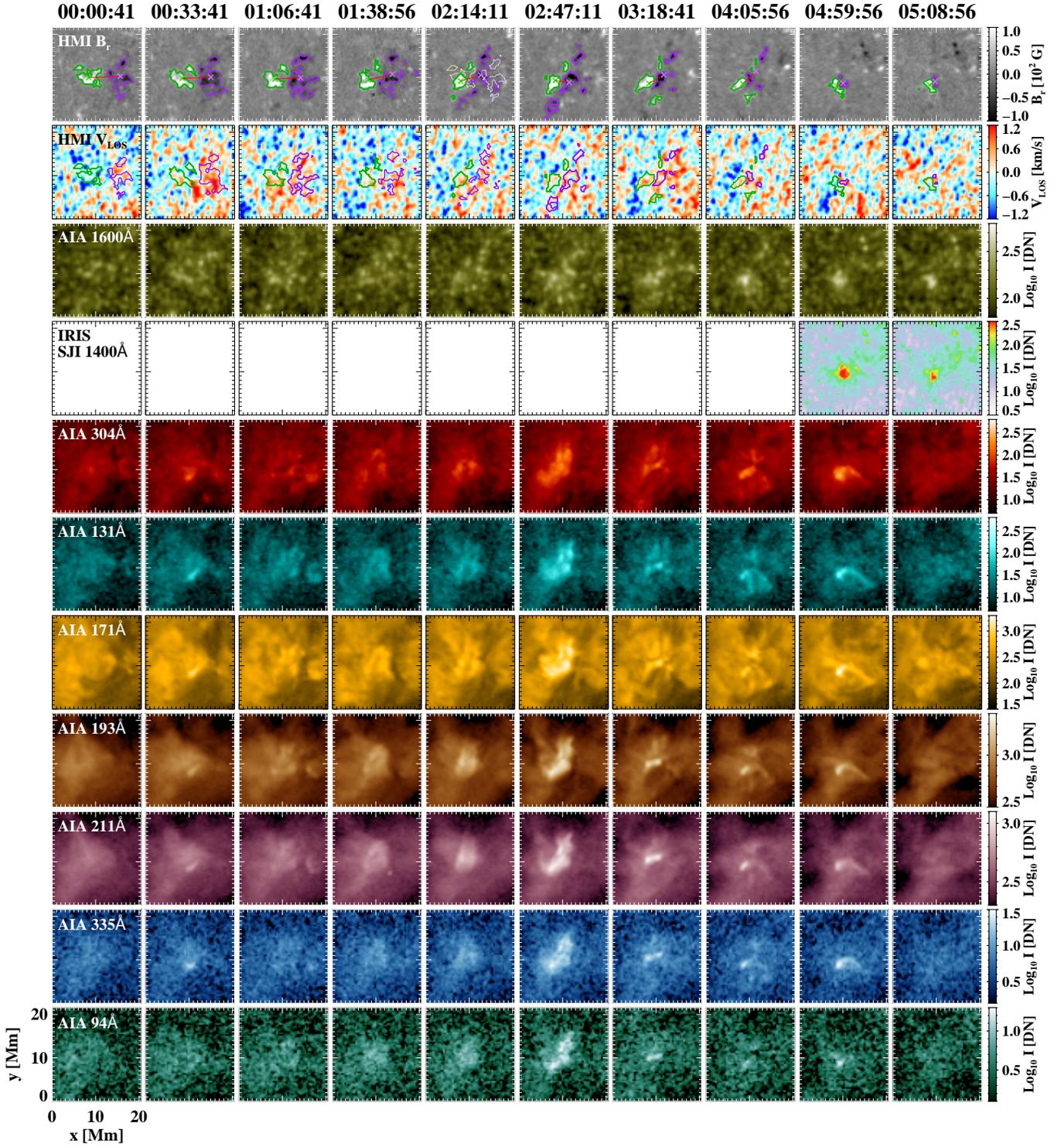}
\caption{A sequence of images for the target interval of 5.2 hr that show the flux cancellation event: (from top to bottom) HMI $B_\mathrm{r}$, HMI $V_\mathrm{LOS}$, AIA 1600\,\AA, \textit{IRIS} SJI 1400\,\AA, all AIA EUV channels (304, 131, 171, 193, 211, 335, and 94\,\AA).}
\label{fig:img_seq}
\end{figure}

We first examine the morphological structure and overall dynamics of the two main magnetic patches of opposite polarities over the target interval, based on a sequence of $B_\mathrm{r}$ images (row 1 in Figure~\ref{fig:img_seq}). At the beginning, the CoM of the leading negative patch is 3.2\,Mm apart from that of the following positive patch. The separation distance between the CoM positions of the leading and following patches continues to decrease throughout the target interval; i.e., the two patches consistently show a converging motion. We also find that at first the line connecting the two CoM positions is tilted $\sim$0{\degr} with respect to the east-–west direction. The tilt angle then gradually increases counterclockwise as the leading/following patches on the photosphere consistently move northward/southward, respectively. This rotation motion between the two opposite-polarity patches along the north--south direction is not thought to be caused by the differential rotation of the solar surface, because their directions are perpendicular to each other. The observed tilt angles disobey Joy’s law; in particular, for the latter half of the target interval they differ from that expected by Joy's law angle by more than 90{\degr}.

\begin{figure}[t!]
\centering
\includegraphics[width=1\textwidth]{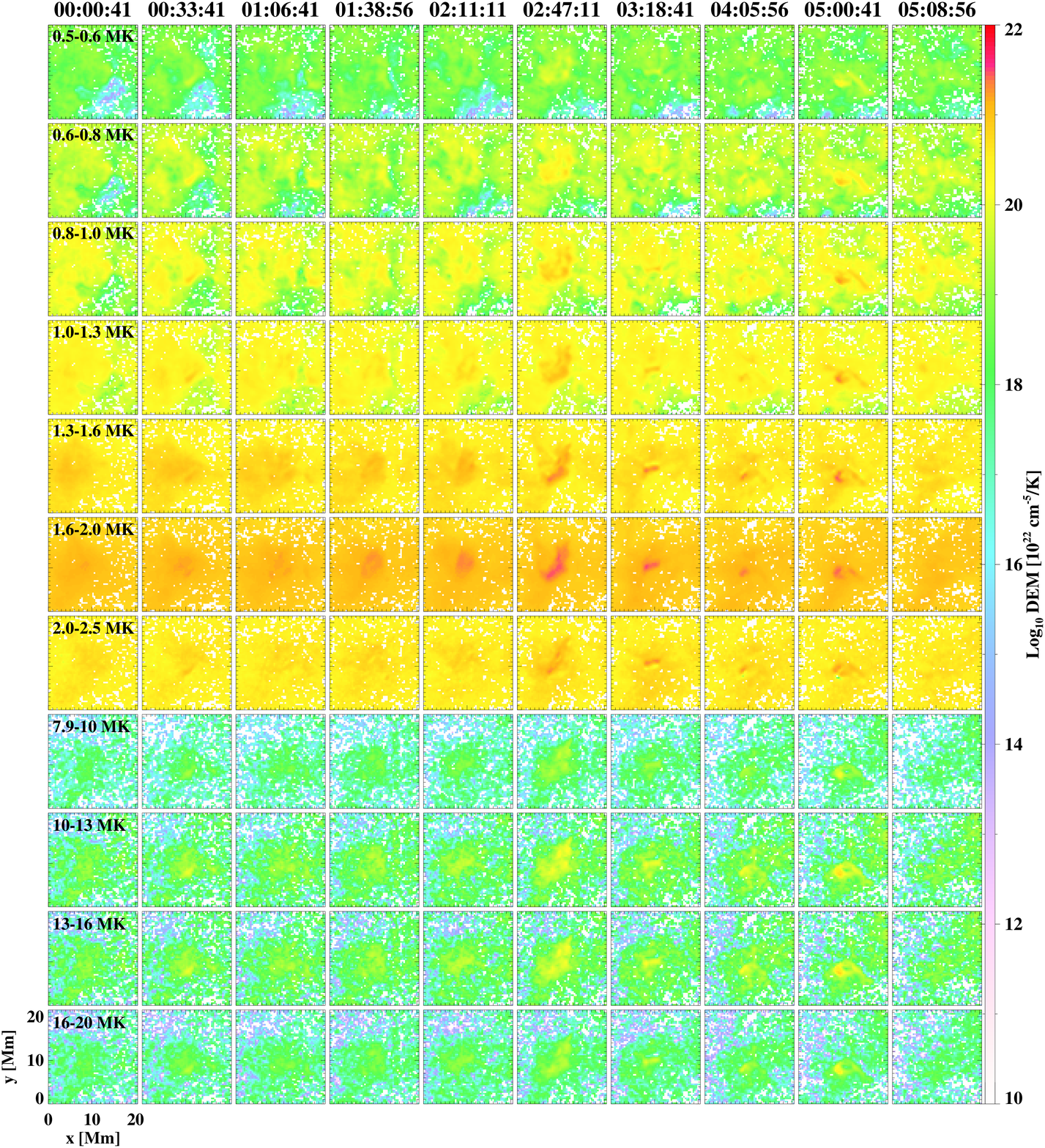}
\caption{A sequence of different emission measure (DEM) maps at different temperature ranges.}
\label{fig:dem}
\end{figure}

In the time-sequence images of $B_\mathrm{r}$ (row 1 in Figure~\ref{fig:img_seq}), it is shown that the pair of the opposite-polarity patches experience a large decrease of magnetic flux in both the positive and negative polarities over the last 1.5 hr of the target interval (i.e., at a rate of 10$^{15}$\,Mx\,s$^{-1}$ as shown in Figure~\ref{fig:patch_outputs}{(b)}). At the same time, the patches decrease in size and eventually disappear. Meanwhile, an emergence of negative-polarity magnetic flux occurs in the middle of the observation (i.e., between $\sim$02:00 to 02:40 TAI; refer to panel (b) of Figure~\ref{fig:patch_outputs}) at the location ($x$,\,$y$)\,$=$\,(8,\,4)\,Mm. A smaller-sized, weaker, positive-polarity counterpart of the emerging negative flux first appears at ($x$,\,$y$)\,$=$\,(13,\,4)\,Mm, which is sufficiently far from the main positive patch (green contour, row 1 in Figure~\ref{fig:img_seq}), so that it is not assigned to the main positive patch until they get close enough at 02:57 TAI to be considered as being merged by the tracking algorithm (cf. Section~\ref{sec:data_analysis}). From HMI LOS Doppler velocity images (row 2, Figure~\ref{fig:img_seq}), signatures of red-shifted Doppler velocities (downflows) are found in most areas of the two patches over the target interval. In multi-wavelength UV and EUV images (rows 3 to 11, from low- to high-temperature channels), we see a couple of impulsive and explosive brightenings localized around the two patches of flux cancellation in the shape of arcades whose two ends are anchored in the main positive and negative magnetic patches on the photosphere (e.g., see the AIA EUV images at 02:47:11 TAI in Figure~\ref{fig:img_seq}). These transient, bright arcade-shaped structures (located around the center of the EUV images) tend to appear as small as $\sim$1\,--\,2\,Mm while the total unsigned flux of the two patches, as well as their CoM separation, decreases during the second half of the target interval. In Figure~\ref{fig:dem}, similar arcade-shaped structures are found in DEM maps over the wide range of T\,$=$\,0.5\,--\,20\,MK. They have relatively large DEM values, compared to those in the background quiet-Sun corona, peaking at T\,$=$\,1.6\,--\,2.0\,MK. Those multi-temperature, arcade structures comprise two distinct populations of electrons at different temperatures: i.e., one in the T\,$=$\,0.5\,--\,2.5\,MK range and the other in the higher temperature range of T\,$=$\,8.0\,--\,20\,MK. Note that the mean coronal temperature is found to be in the range of $\sim$1.4\,--\,1.8\,MK for quiet-Sun regions for years 2010 to 2017 \citep[refer to][]{2017SciA....3E2056M}.

\begin{figure}[t!]
\centering
\includegraphics[width=1\textwidth]{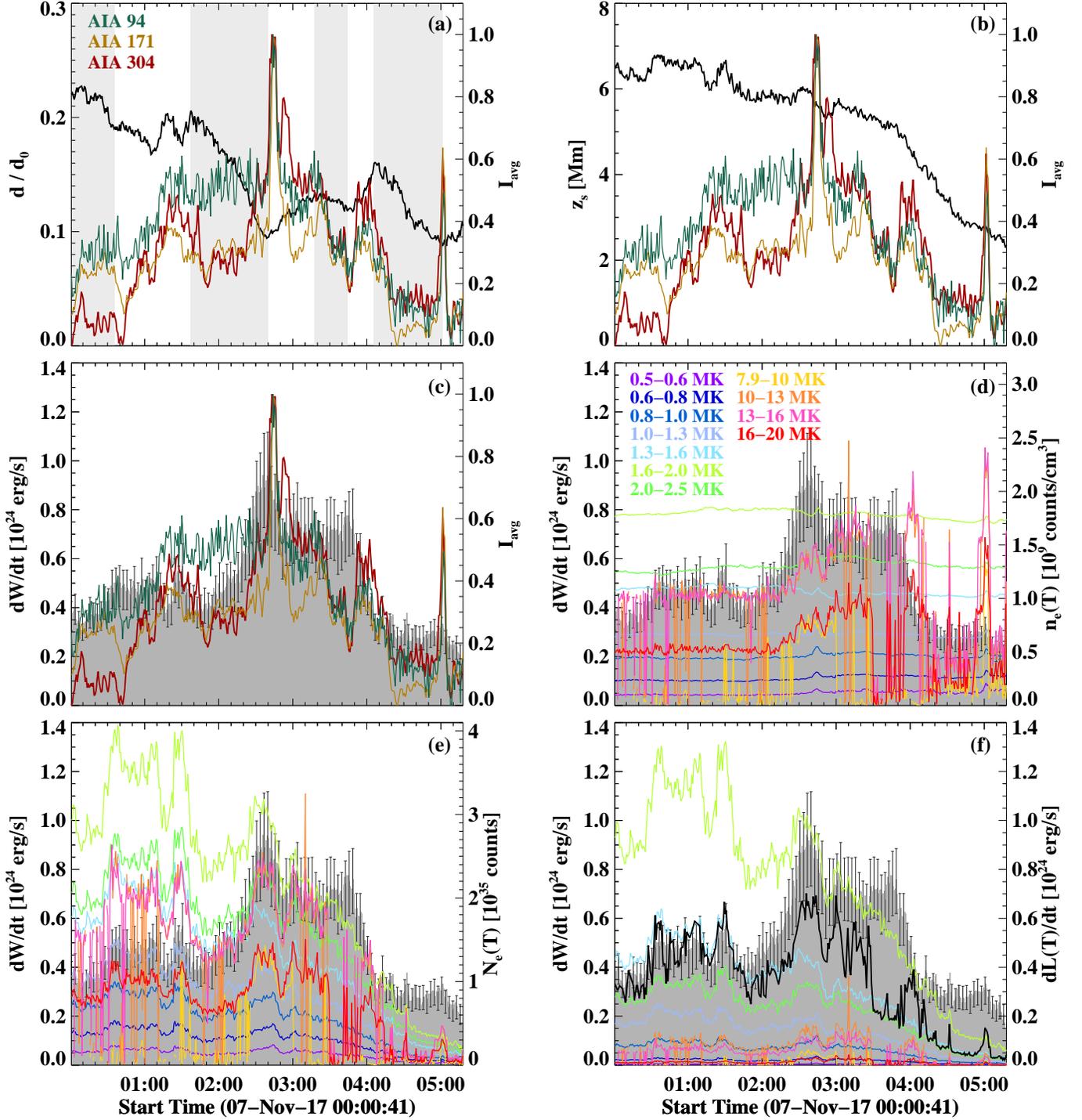}
\caption{Time profiles of (a) the ratio of the separation distance to the interaction distance ($d/d_{0}$, black line) with the normalized average EUV intensities at 94, 171 and 304\,{\AA} ($I_\mathrm{avg}$, colored lines), (b) the height of the reconnecting magnetic separator ($z_{s}$, black line), (c) the total rate of magnetic energy released as heat through reconnection ($\mathrm{d}W/\mathrm{d}t$, gray bars).The vertical error bars on top of the $\mathrm{d}W/\mathrm{d}t$ bar plot indicate upper and lower bounds of $\mathrm{d}W/\mathrm{d}t$. In panels (d--f), $\mathrm{d}W/\mathrm{d}t$ is plotted with the average electron number densities ($n_\mathrm{e}\mathrm{(T)}$), the total electron counts ($N_\mathrm{e}\mathrm{(T)}$), and the radiative energy loss rates ($\mathrm{d}L\mathrm{(T)}/\mathrm{d}t$) at different temperature bins, respectively. The black line with cross symbols in panel (f) represents the sum of the estimated radiative loss rates for electrons at T\,$>$\,1.6\,MK.}
\label{fig:time_profiles}
\end{figure}

In Figure~\ref{fig:time_profiles}, we present how key parameters in the examined reconnection model vary over the target interval of 5.2 hr, specifically in relation to variations in the average EUV intensity over the ROI. Three AIA EUV channels (94, 171 and 304\,{\AA}) are selected here to calculate the average EUV intensity for reference. In the ROI, we first find a long-term (a few hours) variation of a gradual increase and then decrease in the normalized average EUV intensity profiles ($I_\mathrm{avg}$, colored lines in panels (a--c)) from all three different channels over the target interval. On top of the long-term variation, there are numerous impulsive brightenings with two significant peaks in particular at 02:45 and 05:00 TAI. It is also found that the ratio of the separation distance to the interaction distance ($d/d_{0}$, black solid line in panel (a)) shows declining phases (marked with gray bars) at nearly constant rates about 0.5\,--\,1 hr prior to the onset of these EUV brightenings. Moreover, $d/d_{0}$ tends to show relatively large fluctuations and/or persistent increases during the brightenings, compared to the declining phases. The height of the reconnecting magnetic separator ($z_{s}$, panel (b)) is estimated from Equation~(\ref{eq:zs}) to be steadily located at about 6\,Mm until 03:30, but then it decreases until its final estimated height at $\sim$2.5\,Mm.

The total rate of magnetic energy released as heat through reconnection ($\mathrm{d}W/\mathrm{d}t$, gray bars in panel (c) of Figure~\ref{fig:time_profiles}) is estimated to be several $10^{23}$\,erg\,s$^{-1}$, or $10^{5}$\,erg\,cm$^{-2}$\,s$^{-1}$ in the ROI. Upper and lower bounds of $\mathrm{d}W/\mathrm{d}t$ are also derived applying $B_{0}$\,$\pm$\,$\sigma (B_{0})$ to Equation~(\ref{eq:dwdt}), respectively. We find that the estimated magnitude of $\mathrm{d}W/\mathrm{d}t$ corresponds to the energy flux required to heat the quiet-Sun corona \citep[for previous observations, refer to][and references therein]{1977ARA&A..15..363W,2012RSPTA.370.3217P,2017PJAB...93...87S}. In panel (c), we find that $\mathrm{d}W/\mathrm{d}t$ shows similar variations on both short-term (a few to several tens of minutes) and long-term (a few hours) scales as seen in the EUV intensity profiles, sometimes with a time lag of $\sim$5\,--\,10 minutes. This time lag between the magnetic energy release and EUV intensity can be explained by the heating scenario of chromospheric evaporation \citep[first proposed by][]{1968ApJ...153L..59N} in which the rapid downward acceleration of non-thermal particles produced by reconnection, as well as the onset of the consequent evaporation, precedes thermal (soft X-ray and EUV) coronal emissions typically by several minutes \citep{1984ApJ...287..917A,1997ApJ...481..978S,2002A&A...382.1070V,2005ApJ...625.1027K,2016A&A...591A...4M}.

We finally explore how the estimated magnetic energy released as heat by flux cancellation is converted into (or contributes to) radiative energy losses of electrons in the ROI as a function of temperature. For this, the average electron number density ($n_\mathrm{e}\mathrm{(T)}$, shown in panel (d) of Figure~\ref{fig:time_profiles}) over the ROI is first derived at the different ranges of temperature used in the DEM estimations, assuming that the integration distance $d_\mathrm{LOS}$ contributing to the DEM along the LOS is 1\,Mm: i.e., $n_\mathrm{e}\mathrm{(T)}$\,$=$\,$\sqrt{\mathrm{DEM(T)}/d_\mathrm{LOS}}$. The value of $d_\mathrm{LOS}$ used here was chosen, because, in the case of $d_\mathrm{LOS}$\,$=$\,1\,Mm, we can achieve a typical value of $n_\mathrm{e}$(T\,$=$\,1.3\,--\,1.6\,MK)\,=\,$10^{9}$\,cm$^{-3}$ as reported in the quiet-Sun corona. In this calculation of $n_\mathrm{e}\mathrm{(T)}$, we consider only the pixels with the signal-to-noise ratios of the estimated DEM values greater than 5. The total number of electrons ($N_\mathrm{e}\mathrm{(T)}$, panel (e)) is also estimated in the volume of interest $V$\,$=$\,$0.5 \pi S z_{s}$, where $S$ is the total area of both the main positive and negative patches on the photospheric surface. This volume of interest is estimated based on a half-torus geometry with a major (outer) radius of $z_{s}$ and a minor (inner) radius of $\sqrt{0.5 S / \pi}$. It is found that in general temporal variations of the total electron counts at T\,$\geq$\,7.9\,MK follow along with those of $\mathrm{d}W/\mathrm{d}t$, as well as shown in the EUV intensity profiles. The time profiles of $N_\mathrm{e}$(T\,$=$\,1.3\,--\,2.5\,MK) show relatively large values for the first 2 hr of the target interval, when compared to those for the rest interval. The two prominent peaks in the time profiles of $N_\mathrm{e}\mathrm{(T)}$ at 02:45 and 05:00 TAI, are clearly seen only at T\,$\geq$\,7.9\,MK. The total rate of radiative energy loss ($\mathrm{d}L\mathrm{(T)}/\mathrm{d}t$, panel (f) of Figure~\ref{fig:time_profiles}) for a given temperature bin is
\begin{equation}
\frac{\mathrm{d}L\mathrm{(T)}}{\mathrm{d}t} = \int_{V} \tilde{\Lambda}\mathrm{(T)} \,n_\mathrm{e}^{2}\mathrm{(T)}\,\mathrm{d}V,
\label{eq:rad_loss}
\end{equation}
where $\tilde{\Lambda}\mathrm{(T)}$ is the average of the radiative loss function (i.e., emissivity per unit emission measure, adopted from Version 9 of CHIANTI) over the given temperature bin. The short-term dynamic changes in $\mathrm{d}W/\mathrm{d}t$ are shown in variations in $\mathrm{d}L\mathrm{(T)}/\mathrm{d}t$ of electrons at most temperature bins over the target interval, as in $N_\mathrm{e}\mathrm{(T)}$. We also find that values of $\mathrm{d}L\mathrm{(T)}/\mathrm{d}t$ at T\,$=$\,1.0\,--\,2.5\,MK are of the same order of magnitude as $\mathrm{d}W/\mathrm{d}t$, of which overall values are in the range of about $10^{23}$\,--\,$10^{24}$\,erg\,s$^{-1}$. In addition, $\mathrm{d}W/\mathrm{d}t$ is well correlated with the sum (black line in Figure~\ref{fig:time_profiles}{(f)}) of $\mathrm{d}L\mathrm{(T)}/\mathrm{d}t$ for electrons at higher temperatures (i.e., T\,$\geq$\,2.0\,MK) compared to the mean quiet-Sun coronal temperature of $\sim$1.8\,MK in early 2017 reported by \citet{2017SciA....3E2056M}. This supports the validity of our estimate for $\mathrm{d}W/\mathrm{d}t$, in the respect that it can be used to reproduce the observed coronal heating and radiative energy loss over the target interval.

\section{Summary and Conclusions} \label{sec:discussion}
In this paper we have investigated photospheric magnetic field and multi-wavelength UV/EUV observations of a small-scale magnetic flux cancellation event in a quiet-Sun internetwork region of interest (ROI) over a target interval of 5.2 hr. Specifically, we focused on how much the total rate of magnetic energy released as heat ($\mathrm{d}W/\mathrm{d}t$) at the ROI can contribute to the heat requirement of the coronal plasma therein, and also how well temporal variations in $\mathrm{d}W/\mathrm{d}t$ are correlated with those in the total rate of radiative energy loss ($\mathrm{d}L\mathrm{(T)}/\mathrm{d}t$) of electrons at different temperature ranges. To answer the questions, $\mathrm{d}W/\mathrm{d}t$ was estimated using the analytic reconnection model of \citet{2018ApJ...862L..24P} in which key parameters can be derived from the observed photospheric magnetic field of the cancellation event. The regularized inversion method by \citet{2012A&A...539A.146H} was applied to produce the differential emission measure (DEM) maps and eventually $\mathrm{d}L\mathrm{(T)}/\mathrm{d}t$ at the ROI, as a function of temperature, from the multi-channel EUV observations. Our main findings are summarized as follows:
\begin{enumerate}
  \item The observed flux cancellation event involves a pair of opposite-polarity magnetic flux patches on the photosphere that show a converging motion and red-shifted Doppler velocities (downflows) over the target interval, together with an accompanying decrease in magnetic flux of both polarities at a rate of 10$^{15}$\,Mx\,s$^{-1}$ for the last 1.5 hr of the target interval.
  \item Several impulsive EUV brightenings are observed in the form of arcades with their two footpoints anchored in these two magnetic patches.  
  \item The EUV bright arcade-shaped structures are visible in the DEM maps with relatively large DEM values, compared to the background quiet-Sun corona, peaking at T\,$=$\,1.6\,--\,2.0\,MK but with contribution over the broad range of T\,$=$\,0.5\,--\,20\,MK.
  \item The estimated $\mathrm{d}W/\mathrm{d}t$ at the ROI ranges between $2\times10^{23}$ to $1\times10^{24}$\,erg\,s$^{-1}$ over the target interval. Based on the total area of the opposite-polarity patches, the energy flux is determined to be $\sim$10$^{5}$\,erg\,cm$^{-2}$\,s$^{-1}$, which corresponds to the same order of magnitude as previously reported in observations of the quiet-Sun corona.
  \item Both short-term dynamic variations and long-term gradual trends in the 5.2 hr profile of $\mathrm{d}W/\mathrm{d}t$ are clearly shown in $\mathrm{d}L\mathrm{(T)}/\mathrm{d}t$ of electrons at T\,$\geq$\,2.0\,MK.
\end{enumerate}

All these observational findings lend support to the \citet{2018ApJ...862L..24P} model in which reconnection driven by converging and cancelling magnetic patches of opposite polarities on the photosphere can provide sufficient energy to heat the Sun's upper atmosphere (i.e., chromosphere, transition region, corona). Recent high-resolution observations of quiet-Sun magnetic fields are also favorable to the scenario for coronal heating by magnetic flux cancellation, in the context that a large amount of flux cancellation occurs frequently (i.e., at a rate of 10$^{15}$\,Mx\,s$^{-1}$) between dynamic, small-scale magnetic patches on the photosphere \citep[e.g.,][]{2017ApJS..229....4C}. The short-term variations seen in the 5.2 hr profile of $\mathrm{d}W/\mathrm{d}t$ have timescales of a few to several tens of minutes, comparable to the waiting time between successive nanoflares as suggested by hydrodynamic simulations of impulsive coronal heating \citep[e.g.,][]{2014ApJ...784...49C,2016ApJ...833..217B}. It should be noted, however, that flux cancellation events such as the one examined here should be distributed all over the solar surface with a variety of temporal and spatial scales, and they ought to occur at suitable rates to contribute the required energy dissipation rate, or a significant portion thereof, in order to maintain the coronal temperature of the quiet Sun. In simulations by \citet{2019ApJ...872...32S,2020ApJ...891...52S}, distinct outflows/jets were found to occur at different temperatures at reconnection sites during flux cancellation. However, in the small-scale flux cancellation event of this study, no clear morphological or kinematic signature of outflows/jets was found over the target interval amongst the AIA UV/EUV images, or the \textit{IRIS} 1400\,{\AA} SJIs. We note that such narrow sub-arcsecond features as implied by the simulations are difficult to detect at the spatial resolution of AIA and \textit{IRIS}. \textit{IRIS} spectroscopic observations of small-scale flux cancellation events may help identify cancellation-driven outflows/jets and their evolution through temporal analysis of spectral properties, but this is beyond the scope of the present study.

The analytic reconnection model of \citet{2018ApJ...862L..24P} presented in Section~\ref{sec:model} is constructed for a particular case of two equal flux sources of opposite polarities aligned with an overlying uniform horizontal magnetic field. In reality, however, the quiet-Sun magnetic field is more complicated so that the model may need to be improved considering various magnetic field structures and their evolution. The cancellation-based heating model also needs to be further examined with more flux cancellation events observed in both quiet-Sun and active regions as well as in data-driven (meaning realistic) simulations of the solar atmosphere. Moreover, statistical properties of flux cancellation events, such as a space-frequency distribution of estimated magnetic energy released as heat at flux cancellation regions, will help us to examine whether this cancellation-based heating scenario accounts for several aspects of the long-standing coronal heating problem, and also to evaluate how significant the coronal heating by flux cancellation is compared to the other competing heating mechanisms.

\acknowledgments
The author would like to thank an anonymous referee for thoughtful comments, and K. D. Leka, Kanya Kusano, Graham Barnes, Karin Dissauer, and Manolis K. Georgoulis for their valuable comments and suggestions. The data used in this work are courtesy of the NASA/\textit{SDO} and the AIA and HMI science teams, as well as the NASA/\textit{IRIS} team. \textit{IRIS} is a NASA small explorer mission developed and operated by LMSAL with mission operations executed at NASA Ames Research Center and major contributions to downlink communications funded by ESA and the Norwegian Space Centre. This research has made extensive use of the NASA's Astrophysics Data System (ADS) as well as the computer system of the Center for Integrated Data Science (CIDAS), Institute for Space-Earth Environmental Research (ISEE), Nagoya University. This work was partially supported by MEXT/JSPS KAKENHI Grant No.~JP15H05814.
\vspace{5mm}
\facilities{IRIS (SJI), SDO (AIA, HMI)}

\end{document}